\documentclass{aastex}
\usepackage{spr-astr-addons}
\usepackage{url}

\RequirePackage{color}

\begin{document}

\title{Statefinder Parameter for Varying $G$ in Three Fluid System}

\shorttitle{Statefinder Parameter for Varying $G$ in Three Fluid
System} \shortauthors{Hussain et al.}

\author{Mubasher Jamil\altaffilmark{1}}
\and
\author{Muhammad Raza\altaffilmark{1}}
\and
\author{Ujjal Debnath\altaffilmark{2}}

\altaffiltext{1}{Center for Advanced Mathematics and Physics,
National University of Sciences and Technology, Islamabad, Pakistan.
Email: mjamil@camp.nust.edu.pk, jamil.camp@gmail.com;
mreza06@gmail.com, mraza@camp.nust.edu.pk}

\altaffiltext{2}{Department of Mathematics, Bengal Engineering and
Science University, Shibpur, Howrah-711 103, India. Email:
ujjaldebnath@yahoo.com , ujjal@iucaa.ernet.in}

\begin{abstract}

In this work, we have considered variable $G$ in flat FRW universe
filled with the mixture of dark energy, dark matter and radiation.
If there is no interaction between the three fluids, the
deceleration parameter and statefinder parameters have been
calculated in terms of dimensionless density parameters which can be
fixed by observational data. Also the interaction between three
fluids has been analyzed due to constant $G$. The statefinder
parameters also calculated in two cases: pressure is constant and
pressure is variable.

\end{abstract}

\keywords{ }

\section{Introduction}

The present acceleration of the universe as favored by the
Supernovae type Ia data can be explained by some exotic matter
dominated in the present universe which violates the strong energy
condition is termed as dark energy \cite{Perlm, Riess}. This dark
energy has the property that it has positive energy and sufficient
negative pressure \cite{Cald1, Cald2}. Dark energy occupies about
73\% of the energy of our Universe, while dark matter about 23\% and
the usual baryonic matter 4\%. There are different candidates obey
the property of dark energy given by $-$ quintessence \cite{Peebles,
Cald3}, k-essence \cite{Arm}, tachyon \cite{Sen}, phantom
\cite{Cald1}, ghost condensate \cite{Ark, Piazza}, quintom
\cite{Feng, Guo}, brane world models \cite{Sahni} and Chaplygin gas
models \cite{Kamen}.

Einstein's field equations have two parameters $-$ the Newton's
gravitational constant and the cosmological constant. The Newton's
gravitational constant $G$ plays the role of a coupling between
geometry and matter in the Einstein field equations. In an evolving
universe, it appears natural to look at this ``constant" as a
function of time. Dirac \cite{Dirac} proposed for the first time the
idea of a variable $G$ on certain physical grounds. He has shown
that $G\propto t^{-1}$, but his model ran in some difficulties. Some
authors \cite{Rahaman, Mass} have shown that $G$ is an increasing
function of time. Many other extensions of Einstein's theory with
time dependent $G$ have also been proposed in order to achieve a
possible unification of gravitation and elementary particle physics
or to incorporate Mach's principle in general relativity
\cite{Hoyle1, Hoyle2, Brans}. Canuto and Narlikar \cite{Canuto} have
shown that the $G$-varying cosmology is consistent with whatever
cosmological observations presently available. According to Dirac's
large numbers hypothesis, $\dot{G}/G\sim$ Hubble rate $H $
\cite{Melnikov}. Observations of Hulse-Taylor binary pulsar B1913+16
gives the estimate $0<\dot{G}/G\sim2\pm4\times10^{-12}yr^{-1}$
\cite{Kogan} and helioseismological data gives the bound
$0<\dot{G}/G\sim1.6\times 10^{-12}yr^{-1}$ \cite{Guenther, Ray}.
Several works on variable $G $ have been studied in last few decades
\cite{Jamil1, Krori, Jamil2, Arbab1, Arbab2, Arbab3, Arbab4}.

In this work, we have considered the universe is filled with
radiation, dark matter and dark energy with and without
interactions. We now follow the method of the ref.\cite{Jamil2} in
three fluid system. The dimensionless density parameters and
statefinder parameters have been calculated due to variable $G$ in
three fluid system of flat FRW universe.

\section{Three Fluids with Varying Gravitational Constant}

The isotropic, homogeneous and flat FRW model of the universe is
described by the line element
\begin{equation}
ds^{2}=dt^{2}-a^{2}(t)\left[ dr^{2}+r^{2}(d\theta^{2}+\sin^{2}\theta
d\phi^{2})\right],
\end{equation}
where $a(t)$ is the scale factor. The energy-momentum tensor is
given by
\begin{equation}
T_{\mu\nu}=(\rho+p)u_{\mu}u_{\nu}-pg_{\mu\nu},
\end{equation}
where $u_{\mu}$ is the four velocity satisfying $u^{\mu}u_{\mu}=1$.
Here $\rho$ and $p$ are respectively the energy density and pressure
of the fluid in the universe.\newline The Einstein's field equations
are given by
\begin{equation}
R_{\mu\nu}-\frac{1}{2}g_{\mu\nu}R=8\pi G(t)T_{\mu\nu},
\end{equation}
where $R_{\mu\nu}$, $g_{\mu\nu}$ and $R$ are Ricci tensor, metric
tensor and Ricci scalar respectively. Here we consider gravitational
constant $G$ as a function of cosmic time $t$. Now assume that the
universe is filled with the mixture of three fluids (radiation, dark
matter and dark energy), so from the equations (1), (2) and (3) we
have the Einstein's field equations as
\begin{equation}
H^{2}=\frac{8\pi G(t)}{3}\rho,
\end{equation}
and
\begin{equation}
\dot{H}=-4\pi G(t)(\rho+p).
\end{equation}
Also the conservation equation is given by
\begin{equation}
\dot{\rho}+3H(\rho+p)=0,
\end{equation}
where, $\rho=\rho_{x}+\rho_{m}+\rho_{r}$ and $p=p_{x}+p_{m}+p_{r}$
are the total energy density and pressure of the fluid in the
universe. Here, $\rho_{x},\rho_{m}$ and $\rho_{r}$ are the energy
densities of the dark energy, dark matter and radiation fluids
respectively. For dark matter, we choose negligible pressure i.e.,
$p_{m}=0$. Also $p_{x}$ and $p_{r}$ are the pressures of dark energy
and radiation respectively. Now the equation of state for radiation
is given as $p_{r}=\frac{1}{3}\rho_{r}$. Now assume that the
equation of state for dark energy is $p_{x}=w\rho_{x}$. Next we
study the non-interacting and interacting situations.

\section{Non-Interacting case}

If there is no interaction between dark energy, dark matter and
radiation then from equation (6), we can write the continuity
equations for dark matter, dark energy and radiation as
\begin{align}
\overset{.}{\rho}_{m}+3H\rho_{m} &  =0,\\
\overset{.}{\rho}_{x}+3H\left(  1+w\right)  \rho_{x} &  =0,\\
\overset{.}{\rho}_{r}+4H\rho_{r} &  =0.
\end{align}
Now we define the dimensionless density parameters in the form
\begin{equation}
\Omega=\frac{\rho}{\rho_{cr}}~,~\Omega_{x}=\frac{\rho_{x}}{\rho_{cr}}%
~,~\Omega_{m}=\frac{\rho_{m}}{\rho_{cr}}~,~\Omega_{r}=\frac{\rho_{r}}%
{\rho_{cr}}~
\end{equation}
where $\rho_{cr}$ is the critical density, so we obtain
\begin{equation}
\Omega=\Omega_{x}+\Omega_{m}+\Omega_{r}.%
\end{equation}
The deceleration parameter $q=-\frac{\overset{..}{a}}{aH^{2}}$ can
be expressed as in terms of density parameters in the following
\begin{equation}
q=\frac{1}{2}[\left(  1+3w\right)
\Omega_{x}+\Omega_{m}+\Omega_{r}].
\end{equation}
From above we obtain
\begin{equation}
\overset{.}{q}=\frac{1}{2}[\left(  1+3w\right)  \overset{.}{\Omega}%
_{x}+3\overset{.}{w}\Omega_{x}+\overset{.}{\Omega}_{m}+\overset{.}{\Omega}%
_{r}].
\end{equation}
From (9), we have
\begin{equation}
\overset{.}{\Omega}=\frac{\overset{.}{\rho}}{\rho_{cr}}-\frac{\rho}{\rho
_{cr}^{2}}\overset{.}{\rho}_{cr},%
\end{equation}
where
\begin{equation}
\overset{.}{\rho}_{cr}=\rho_{cr}\left(  2\frac{\overset{.}{H}}{H}%
-\frac{\overset{.}{G}}{G}\right).
\end{equation}
Furthermore,
\begin{equation}
\overset{.}{H}=-H^{2}\left(  1+q\right).
\end{equation}
So from (14) we have
\begin{equation}
\overset{.}{\rho}_{cr}=-H\rho_{cr}\left(  2\left(  1+q\right)
+\bigtriangleup G\right),
\end{equation}
where $\Delta_{G}\equiv G^{\prime}/G$, $\dot{G}=HG^{\prime}$. Now
from (13) we get
\begin{equation}
\overset{.}{\Omega}=\frac{\overset{.}{\rho}}{\rho_{cr}}+\Omega
H\left[ 2\left(  1+q\right)  +\bigtriangleup G\right].
\end{equation}
From equations (7) to (10), we yield
\begin{equation}
{\dot{\Omega}}_{m}=\Omega_{m}H\left[  -1+2q+\bigtriangleup G\right]
,
\end{equation}%
\begin{equation}
{\dot{\Omega}}_{x}=\Omega_{x}H\left[  -1-3w+2q+\bigtriangleup
G\right],
\end{equation}
and
\begin{equation}
{\dot{\Omega}}_{r}=\Omega_{r}H\left[  -2+2q+\bigtriangleup G\right].
\end{equation}
Substituting equations (19) - (21) in equation (13) and after
simplifying, we get
\begin{align}
\overset{.}{q}  & =\left.  \frac{1}{2}\Omega_{m}H\left(
-1+2q+\bigtriangleup
G\right)  +\Omega_{x}H\left\{  \left(  1+3w\right)  +\frac{3}{2}\overset{.}%
{w}\right\}  \right. \nonumber\\
& \left.  \left(  -1-3w+2q+\bigtriangleup G\right)
+\Omega_{r}H\left( -2+2q+\bigtriangleup G\right)  ]\right.
\end{align}

We also determine the dimensionless pair of cosmological diagnostic
pair $\{r,s\}$ dubbed as statefinder parameters introduced by
\cite{Sahni1}. The two parameters have a great geometrical
significance since they are derived from the cosmic scale factor
alone. Also this pair generalizes the well known geometrical
parameters like the Hubble parameter and the deceleration parameter.
The parameter $r$ forms the next step in the hierarchy of
geometrical cosmological parameters $H$ and $q$.

The diagnostic pair has been used for holographic dark energy model
\cite{setare,setare1,setare2,setare3,setare4}, Chaplygin gas
\cite{jamil3,jamil4}.

The parameters are given by
\begin{equation}
r=\frac{\overset{...}{a}}{aH^{3}}~~\text{and}~~s=\frac{r-1}{3(q-1/2)}.%
\end{equation}
For varying gravitational constant $G$, the definition of $s$ can be
generalized to \cite{Jamil2}
\begin{equation}
s=\frac{r-\Omega}{3\left(  q-\frac{\Omega}{2}\right)  }.%
\end{equation}
From (23), we can write $r$ in terms of $H$ and $q$ as
\begin{equation}
r=2q^{2}+q-\frac{\overset{.}{q}}{H}.%
\end{equation}
Using (12) and (22), we can write equation (25) in the form
\begin{align}
r &  =\left.  \frac{1}{2}[\left(  1+3w\right)  \Omega_{x}+\Omega_{m}%
+\Omega_{r}]^{2}\right.  \nonumber\\
&  \left.  +\frac{1}{2}[\left(  1+3w\right)
\Omega_{x}+\Omega_{m}+\Omega
_{r}]\right.  \nonumber\\
&  \left.  -\frac{1}{2}\Omega_{m}\left(  -1+2q+\bigtriangleup
G\right)
\right.  \nonumber\\
&  \left.  +\Omega_{x}\left\{  \left(  1+3w\right)  +\frac{3}{2}\overset{.}%
{w}\right\}  \times\right.  \nonumber\\
&  \left.  \left(  -1-3w+2q+\bigtriangleup G\right)
-\Omega_{r}\left( -2+2q+\bigtriangleup G\right)  ]\right.
\end{align}
Substituting this in equation (24), we get
\begin{align}
s &  =\frac{1}{3w\Omega_{x}}\left\{  [\left(  1+3w\right)  \Omega_{x}%
+\Omega_{m}+\Omega_{r}]^{2}\right.  \nonumber\\
&  \left.  +[\left(  -1+3w\right)
\Omega_{x}+\Omega_{r}]-\Omega_{m}\left(
2q+\bigtriangleup G\right)  \right.  \nonumber\\
&  \left.  +2\Omega_{x}\left\{  \left(  1+3w\right)
+\frac{3}{2}\overset {.}{w}\right\}  \times\left(
-1-3w+2q+\bigtriangleup G\right)  \right.
\nonumber\\
&  \left.  -2\Omega_{r}\left(  -1+2q+\bigtriangleup G\right)
]\right\}
\end{align}

\bigskip

\section{Interacting Case}

Interacting models where the dark energy weakly interacts with the
dark matter have also been studied to explain the evolution of the
Universe. This models describe an energy flow between the
components. To obtain a suitable evolution of the Universe an
interaction is often assumed such that the decay rate should be
proportional to the present value of the Hubble parameter for good
fit to the expansion history of the Universe as determined by the
Supernovae and CMB data \cite{Berger}. These kind of models describe
an energy flow between the components so that no components are
conserved separately. First, we assume that the dark matter
component is interacting with dark energy component, so the
continuity equations of dark matter and dark energy are
\begin{align}
\overset{.}{\rho}_{m}+3H\rho_{m}  &  =Q,\\
\overset{.}{\rho}_{x}+3H\left(  1+w\right)  \rho_{x}  &  =-Q^{\prime},%
\end{align}
where $Q$ and $Q^{\prime}$ in order to include the scenario in which
the mutual interaction between the two principal components of the
universe leads to some loss in other forms of cosmic constituents.
In this case, we have assumed $Q\ne Q^{\prime}$, so from (6), we
have the continuity equation for radiation fluid as \cite{Cruz,
Debnath,jamil2,rahaman2}
\begin{align}
\overset{.}{\rho}_{r}+4H\rho_{r}  &  =Q'-Q.
\end{align}
If $Q<Q^{\prime}$, then part of the dark energy density decays into
dark matter and the rest in the radiation fluid. But if
$Q>Q^{\prime}$, then dark matter receives energy from dark energy
and from radiation. We are taking about in this case that dark
energy decay into dark matter (or vice versa, depending on the sign
of $Q$) and radiation. Assume, the interaction terms $Q $ and
$Q^{\prime}$ are $Q=3\Pi_{1}H$ and $Q^{\prime}=3\Pi_{2}H$ which
measure the strength of interactions where $\Pi_{1}$ and $\Pi_{2}$
have the dimension of density. Now assume that $G$ is constant. So
Differentiating equation (5), we have
\begin{equation}
\ddot{H}=-4\pi G\left(  \dot{\rho}+\dot{p}\right).
\end{equation}
Also the deceleration parameter $q$ can be expressed as
\begin{equation}
q=-\frac{1}{2}-\frac{3}{2}\frac{p}{\rho}.%
\end{equation}

${\bullet}$ \textbf{Case-I:} when $\dot{p}=0$. Equation (31) reduces
to
\begin{eqnarray}
\ddot{H}&=&-4\pi GH\Big[  -3\rho_{m}-3\Big( 1-\dot{w}-w^{2}\Big)
\rho_{x}\nonumber\\&&-w\frac{Q^{\prime}}{H}\rho_{x}-4\rho_{r}\Big].
\end{eqnarray}
Dividing by $H^{2}$, we yield
\begin{equation}
\frac{\ddot{H}}{H^{3}}=9-9\frac{\left\{  \left(  \dot{w}+w\Pi_{2}%
+w^{2}\right)  \rho_{x}+\left(  -\frac{7}{3}\right)  \rho_{r}\right\}  }{\rho}.%
\end{equation}

Now equation (25) can be written as
\begin{equation}
r=\frac{\ddot{H}}{H^{3}}-3q+2.
\end{equation}
Using (34), the above equation becomes
\begin{equation}
r=\frac{25}{2}+\frac{9p}{4\rho}-9\frac{\left(
\dot{w}+w\Pi_{2}+w^{2}\right).
\rho_{x}}{\rho}%
\end{equation}
Also from (23), we obtain the expression for $s$ as
\begin{equation}
s=\frac{\frac{23}{2}+\frac{9p-36\left(  \dot{w}+w\Pi_{2}+w
^{2}\right)
\rho_{x}}{4\rho}}{3\left(  -1-\frac{3}{2}\frac{p}{\rho}\right)  }.%
\end{equation}

${\bullet}$ \textbf{Case-II:} when \ $\dot{p}\neq0$ then for $\ p=p_{x}%
+p_{r}=w\rho_{x}+\frac{1}{3}\rho_{r}$ we get (from (31))
\begin{align}
\frac{\ddot{H}}{H} &  =-4\pi G\left\{  -3\rho_{m}+\left(  -3-6w-3w^{2}%
+\frac{1}{H}\dot{w}\right)  \rho_{x}\right.  \nonumber\\
&  \left.  -\frac{16}{3}\rho_{r}-\left(  3w-1\right)
\Pi_{2}-\Pi_{1}\right\}
\times\nonumber\\
&  \left\{  \left(  1-\dot{w}\right)
-w\frac{Q^{\prime}}{H}\rho_{x}-4\rho _{r}\right\}.
\end{align}
Dividing by $H^{2}$ we obtain
\begin{align}
\frac{\ddot{H}}{H^{3}} &  =\frac{3}{2}\frac{1}{\rho}\left\{  -3\rho
_{m}+\left(  -3-6w-3w^{2}+\frac{1}{H}\dot{w}\right)  \rho_{x}\right.
\nonumber\\
&  \left.  -\frac{16}{3}\rho_{r}-\left(  3w-1\right)
\Pi_{2}-\Pi_{1}\right\}
\times\nonumber\\
&  \left\{  \left(  1-\dot{w}\right)
-w\frac{Q^{\prime}}{H}\rho_{x}-4\rho _{r}\right\}.
\end{align}

In this case, the expressions of $r$ and $s$ become
\begin{equation}
r=\frac{7}{2}-\frac{9}{2}\frac{\rho_{m}}{\rho}+\left(  \frac{3}{2}D_{1}%
+\frac{9}{2}w\right)
\frac{\rho_{x}}{\rho}-\frac{13}{2}\frac{\rho_{r}}{\rho
}-\frac{3}{2}\frac{D_{2}}{\rho},%
\end{equation}
and
\begin{equation}
s=\frac{\frac{5}{2}-\frac{9}{2}\frac{\rho_{m}}{\rho}+\left(  \frac{3}{2}%
D_{1}+\frac{9}{2}w\right)  \frac{\rho_{x}}{\rho}-\frac{13}{2}\frac{\rho_{r}%
}{\rho}-\frac{3}{2}\frac{D_{2}}{\rho}}{3\left(
-1-\frac{3}{2}\frac{p}{\rho
}\right)  },%
\end{equation}
where
\begin{align*}
D_{1} &  =-3-6w-3w^{2}+\frac{1}{H}\dot{w},\\
D_{2} &  =-\left(  3w-1\right)  \Pi_{2}-\Pi_{1}.%
\end{align*}

\section{Discussions}

In this work, we have considered variable $G$ in flat FRW universe
filled with the mixture of dark energy, dark matter and radiation.
If there is no interaction between the three fluids, the
deceleration parameter and statefinder parameters have been
calculated in terms of dimensionless density parameters which can be
fixed by observational data. Also the interaction between three
fluids has been analyzed due to constant $G$. If $Q<Q^{\prime}$,
then part of the dark energy density decays into dark matter and the
rest in the radiation fluid. But if $Q>Q^{\prime}$, then dark matter
receives energy from dark energy and from radiation. We are taking
about in this case that dark energy decay into dark matter (or vice
versa, depending on the sign of $Q$) and radiation. The statefinder
parameters also calculated in two cases: pressure is constant and
pressure is variable. In the literature, the diagnostic pair has
been calculated for the dark energy interacting with dark matter
\cite{jamil2}. The interaction of these two dark fluids with a third
component has been studied in literature, and hence motivated by
this, we calculated the statefinder parameters for the triple fluid
interacting model. This model will be useful for investigating
further the triple coincidence problem \cite{Ark}.

\end{document}